# Two-dimensional spin-valley-coupled Dirac semimetals in functionalized SbAs monolayers

Zhifeng Liu,[†‡] Wangxiang Feng,[§l] Hongli Xin,[†] Yinlu Gao,[⊥]  Pengfei Liu,[‡#]
Yugui Yao,[§] Hongming Weng,[£¶] Jijun Zhao[‡⊥*]

[†]*School of Physical Science and Technology, Inner Mongolia University, Hohhot 010021, China*

[‡]*Beijing Computational Science Research Center, Beijing 100094, China*

[§]*School of Physics, Beijing Institute of Technology, Beijing 100081, China*

[l]*Peter Grünberg Institut and Institute for Advanced Simulation,  Forschungszentrum Jülich and JARA, 52425 Jülich, Germany*

[⊥]*Key Laboratory of Materials Modification by Laser, Ion and Electron Beams (Dalian University of Technology), Ministry of Education, Dalian 116024, China*

[#]*Institute of High Energy Physics, Chinese Academy of Sciences (CAS), Beijing 100049, China*

[£]*Beijing National Laboratory for Condensed Matter Physics and Institute of Physics, Chinese Academy of Sciences, Beijing 100190, China*

[¶]*Songshan Lake Materials Laboratory, Guangdong 523808, China*

[*] Corresponding author. Email: zhaojj@dlut.edu.cn

**ABSTRACT:**

In the presence of spin-orbit coupling (SOC), achieving both spin and valley polarized Dirac state is significant to promote the fantastic integration of Dirac physics, spintronics and valleytronics. Based on *ab initio* calculations, here we demonstrate that a class of *spin-valley-coupled Dirac semimetals* (svc-DSMs) in the functionalized SbAs monolayers (MLs) can host such desired state. Distinguished from the graphene-like 2D Dirac materials, the Dirac cones in svc-DSMs hold giant spin-splitting induced by strong SOC under inversion symmetry breaking. In the 2.3% strained $SbAsH_2$ ML, the Dirac fermions in inequivalent valleys have opposite Berry curvature and spin moment, giving rise to *Dirac spin-valley Hall effect* with constant spin Hall conductivity as well as massless and dissipationless transport. Topological analysis reveals that the svc-DSM emerges at the boundary between trivial and 2D topological insulators, which provides a promising platform for realizing the flexible and controllable tuning among different quantum states.

The rise of graphene[1-3] has inspired significant efforts in searching for other 2D Dirac materials (DMs)[4-10] with linear energy dispersion. As the host of massless Dirac fermions, 2D DMs have become the playground for investigating many quantum relativistic phenomena[11-13] in the emerging field of Dirac physics. Like graphene, without considering spin-orbit coupling (SOC), the Dirac points in these 2D DMs are protected by symmetry. However, the presence of SOC will open a global bulk gap at the Dirac points and introduce the topologically protected gapless edge states.[13-15] Therefore, 2D DMs are formally a quantum spin Hall insulator[13] under time-reversal symmetry [Fig. 1(a)] or quantum anomalous Hall insulator[12] with time-reversal breaking [Fig. 1(b)]. In a pioneering theoretical study, Yong and Kane[16] employed symmetry analysis and a two-site tight-binding model to examine the possibility that the Dirac points can not be gapped by SOC [Fig. 1(c)]. They concluded that nonsymmorphic space group symmetry plays an essential role in protecting the 2D Dirac points against SOC, named as spin-orbit Dirac points (SDPs). Although the first realistic 2D material hosting SDPs was predicted recently[17], its SDPs are not located at the Fermi level and its Dirac dispersion is contaminated by some extraneous non-Dirac bands. Hence, how to achieve a 2D Dirac semimetal hosting SDPs with clean Dirac bands at the Fermi level remains a great challenge.

On the other hand, the discovery of valley-dependent effects in $MoS_2$ monolayer[18,19] without inversion symmetry aroused an upsurge in the field of 2D valleytronics.[20] For hexagonal 2D materials such as graphene and monolayer group-VI transition metal dichalcogenides, the conically shaped valleys at +K and −K corners are inequivalent but related by time-reversal symmetry. Owing to the large separation between these inequivalent valleys in the momentum space, the intervalley scattering is greatly suppressed in pure samples. For this reason, the degree of valley can be used to store and carry information, similar to spin in spintronics.[20,21] It has been confirmed that the breaking of inversion symmetry is a general scheme to generate and detect valley polarization through the valley Hall effect.[21] Interestingly, inversion symmetry breaking together with SOC also lifts spin degeneracy and then leads to valley-contrasting spin splitting, which is the foundation for integration of the spin and valley physics (namely, spin-valleytronics).[18,22] When an in-plane electric field is applied, the charge carriers with opposite valley and spin indexes would gain opposite anomalous transverse velocity, giving rise to simultaneous spin and valley Hall effect.[18,21] Since this effect is induced under time-reversal symmetry, the reversible transverse spin Hall current should be dissipationless,[23,24] which would be an advantageous feature for the next generation ultra-low-power nanodevices. For device applications, the desirable ultra-fast transport requires massless charge carriers derived from the Dirac dispersion. To this end, an intriguing question arises — whether the integration of Dirac physics with spin-valleytronics can be realized in a realistic 2D material?

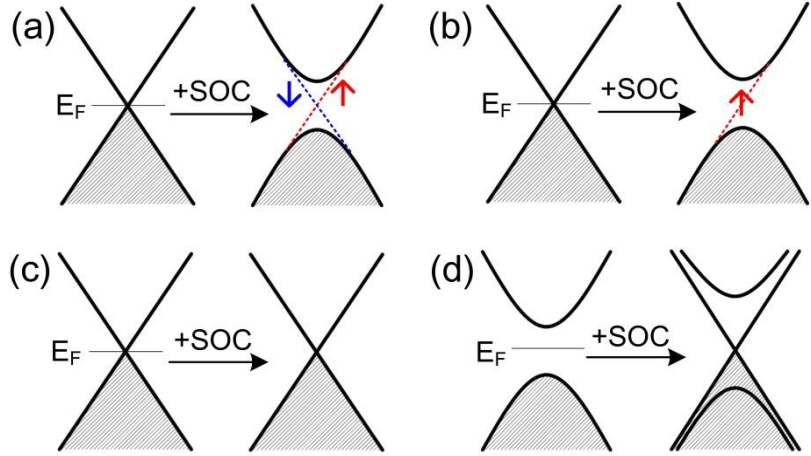

FIG. 1. Schematic electronic band structures without SOC (left) and with SOC (right) for different 2D Dirac materials: (a) quantum spin Hall insulator, (b) quantum anomalous Hall insulator, (c) symmetry-protected Dirac semimetal and (d) svc-DSMs. The solid black lines represent 2D bulk states. The dotted lines in (a) and (b) indicate the gapless edge states counter-propagating along the same boundary with opposite spins, and chiral gapless edge state located at the boundary of quantum anomalous Hall insulator, respectively. The red and blue colors denote spin up and spin down, respectively. The shaded areas indicate the occupied states.

In this Letter, we demonstrate that such integration can indeed be realized in a class of 2D spin-valley-coupled Dirac semimetals (svc-DSMs) [Fig. 1(d)], which own the following merits distinguished from the ordinary 2D DMs: (i) their Dirac state is formed in the presence of SOC at inversion asymmetric points in the Brillouin zone (BZ), hence the corresponding Dirac points are intrinsically robust against SOC; (ii) the Dirac bands are spin nondegenerate and exhibit evident spin-splitting at the Dirac points, which are clean in the spin-splitting energy window without contamination by the non-Dirac bands; (iii) massless Dirac fermions in the inequivalent Dirac valleys have opposite Berry curvature and spin moment. With these unique features, the so-called *Dirac spin-valley Hall effect* would occur in a svc-DSM under in-plane electric field, in which ultra-fast and ultra-low-power transport can be achieved simultaneously. Furthermore, since the Dirac dispersion is a 2D bulk state, such novel transport should exist in the whole bulk rather than only the edge of the host material.

This is distinctly different from the quantum spin Hall effect in the conventional 2D topological insulators (TIs). Using *ab initio* calculations, we identify that the functionalized SbAs monolayers [SbAsX$_2$ (X= H, F, Cl, Br and I)] under modest tensile strain are such novel 2D svc-DSMs. Because of their giant spin-splitting and high Fermi velocity, SbAsX$_2$ monolayers (MLs) should be the promising platforms for *Dirac spin-valleytronics* applications at room temperature.

Our *ab initio* calculations based on the density functional theory were performed by the PAW method [25] with a kinetic energy cutoff of 500 eV for the planewave basis, as implemented in the VASP code.[26] The generalized gradient approximation with PBE parameterization[27] was adopted to treat the exchange-correlation interactions. Monkhorst-Pack *k*-point mesh with a grid density of 2π×0.01 Å$^{-1}$ was applied to sample the BZ. All atoms were fully optimized until the change in total energy and the residual force were less than $10^{-6}$ eV and $10^{-3}$ eV/Å, respectively. The hybrid functional HSE06[28] was also employed to obtain more accurate band structures for the svc-DSMs. To explore the topological properties, the maximally localized Wannier functions were constructed by the Wannier90 code,[29] and the edge states were calculated using the iterative Green's function method in the WannierTools package.[30]

Recently, the chemically ordered SbAs compound with layered structure ($R\bar{3}m$) resembling the bulk arsenic and antimony has been synthesized.[31] Subsequent theoretical reports[32-34] revealed that its ML honeycomb buckled structure with inversion asymmetry is an indirect semiconductor, which can transform into a direct semiconductor and even a 2D TI under tensile strain.[34] Although such fascinating ML has not been synthesized in experiment yet, the low cleavage energy, satisfactory stability and strong in-plane stiffness (see Figs. S1 and S2 in the Supporting Information) strongly suggest that a freestanding SbAs-ML should be readily obtained from bulk SbAs crystal via mechanical exfoliation just like graphene.[1]

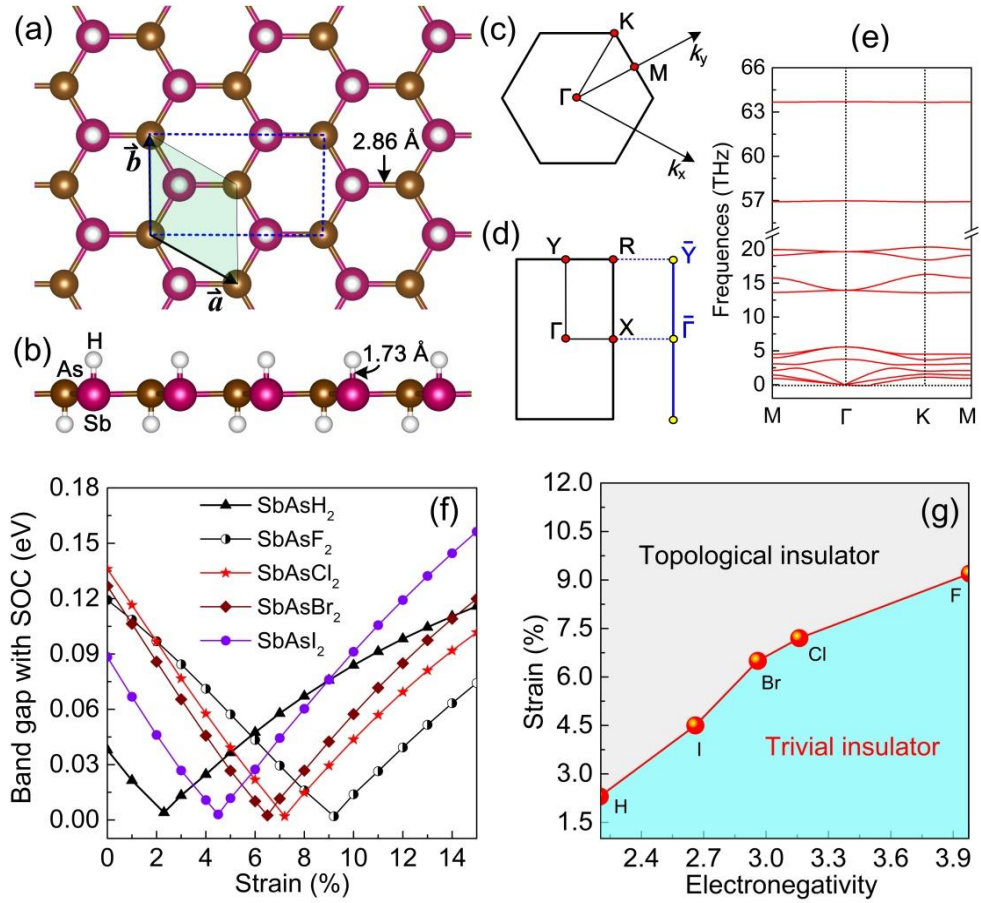

FIG. 2. The (a) top and (b) side views for the equilibrium structure of SbAsH$_2$ ML. The shaded area with Bravais lattice vectors (***a*** and ***b***) marks its primitive cell, while the area with dotted blue lines stands for a redefined lattice (2***a***+***b***, ***b***). 2D BZ together with high symmetry points for (c) primitive cell and (d) redefined lattice (the projected edge BZ for zigzag nanoribbons is also displayed). (e) Phonon dispersion of SbAsH$_2$. (f) The variation of energy band gap with SOC as a function of biaxial tensile strain for SbAsX$_2$ MLs. (g) Phase diagram — critical strain for the transition between topological insulator and trivial insulator versus electronegativity (Pauling scale) of the functional group X.

Owing to the orbital-filtering effect, functionalization is an efficient approach to tailor the chemical, physical and topological properties of 2D materials.[35-39] Once the aforementioned freestanding SbAs ML has been obtained, it can be readily functionalized by X atoms (X = H, F, Cl, Br, I) in laboratory, just like the hydrogenated graphene and fluorinated graphene.[38,39] Upon optimization, the X-decorated SbAs MLs (i.e., SbAsX$_2$) relax to a quasi-planar configuration with the functional groups alternating on both sides [see Fig. 2(b)]. This can be understood by the fact that the X atoms can produce in-plane tensile strain on the SbAs framework, which flattens the pristine buckled structure. As a representative, Figs. 2(a) and 2(b) depict the equilibrium structure of SbAsH$_2$. After hydrogenation, the buckling height decreases from 1.52 Å to 0.06 Å with respect to the pristine SbAs ML, while the lattice constant increases from 3.86 Å to 4.95 Å, and the Sb-As bond length from 2.70 Å to 2.86 Å, respectively. Even so, the trigonal Bravais lattice ($P3m1$) with broken inversion symmetry remains invariant, which can be described by the $C_{3v}$ point group symmetry. The dynamical stability of SbAsX$_2$ was further confirmed by phonon dispersion calculations, showing no imaginary frequency [Fig. 2(c)]. Moreover, the formation energies of SbAsX$_2$ [defined as $E_{form} = (E_{total} - E_{SbAs-ML} - 2\mu_X) / 4$, where $E_{total}$ and $E_{SbAs-ML}$ are the total energies of SbAsX$_2$ and SbAs MLs, respectively, and $\mu_X$ is the chemical potential calculated from gaseous X$_2$ molecule] have also been computed to assess their thermodynamic stability. As listed in Table I, all the calculated $E_{form}$ values of SbAsX$_2$ are distinctly negative, implying exothermic functionalization of X atoms on the pristine SbAs ML.

TABLE I. The lattice constant $a$ (Å), formation energy $E_{form}$ (eV/atom), and band gap (meV) without SOC $E_g$ and with SOC $E_g^{soc}$ for SbAsX$_2$ MLs. The critical tensile strain $\varepsilon$ at which the svc-DSM state emerges. The spin splitting (meV) of LCB ($\Delta_e$) and UVB ($\Delta_h$) at the $K$ point, and Fermi velocity $v_F$ ($\times 10^5$ m s$^{-1}$) for the corresponding svc-DSM state of SbAsX$_2$ MLs. For comparison, the relevant data of pristine SbAs ML are also presented. The data in parentheses are from HSE06 calculations.

|  | $a$ | $E_f$ | $E_g$ | $E_g^{soc}$ | $\varepsilon$ | $\Delta_e$ | $\Delta_h$ | $v_F$ |
|---|---|---|---|---|---|---|---|---|
| SbAsF$_2$ | 4.86 | −1.253 | 357 | 119 | 9.2% | 376 (461) | 185 (274) | 6.24 |
| SbAsCl$_2$ | 4.93 | −0.459 | 434 | 136 | 7.2% | 419 (498) | 186 (283) | 6.44 |
| SbAsBr$_2$ | 4.94 | −0.355 | 433 | 127 | 6.5% | 437 (508) | 196 (289) | 6.59 |
| SbAsI$_2$ | 4.97 | −0.237 | 409 | 88 | 4.5% | 448 (515) | 199 (294) | 7.32 |
| SbAsH$_2$ | 4.95 | −0.977 | 336 | 38 | 2.3% | 421 (475) | 189 (267) | 8.79 |
| SbAs | 3.86 |  | 1470 | 1270 |  |  |  |  |

Without considering SOC, SbAs-ML is an indirect semiconductor (see Fig. S3 in the Supporting Information), while all SbAsX$_2$ MLs are direct semiconductors (Figs. S4 and S5) with largely reduced band gaps (see $E_g$ in Table I) at the $K$ point. Such noticeable change is mainly attributed to the strong coupling between the functional groups and the $p_z$ orbitals of Sb and As atoms, which induces a large local gap at the $\Gamma$ point and two opposite valleys (hole and electron pockets) with a small gap at the K point. With inclusion of SOC, the SbAsX$_2$ MLs remain direct semiconductors (Figs. S4 and S5), but the band gaps are further decreased due to the giant spin-splitting (see $\Delta_e$ and $\Delta_h$ in Table I) resulted from the significant SOC effect in heavy elements (Sb, As) and the missing inversion symmetry. Compared with the elementary group-V MLs decorated with functional groups [e.g., fluorinated As,[33] BiX/SbX (X=H, F, Cl,

Br)[40]], two different aspects should be noted: (i) without SOC the functionalization of binary SbAs ML cannot form gapless Dirac state like that for the elementary MLs due to the broken inversion symmetry in SbAs ML; (ii) with SOC the band gaps of SbAsX$_2$ are greatly reduced rather than widely opened from the Dirac state as in the elementary MLs. These unique features urge us to explore whether the two opposite energy pockets at the K points can be tailored to touch with each other and form the craved svc-DSM state.

There are two critical factors that affect the global bulk gap of SbAsX$_2$. One is the electronegativity of X atoms, which should determine the asymmetric potential gradient in the vertical direction of the basal plane. The other is the strength of SOC that is responsible for the amplitude of spin-splitting $\Delta_e$ and $\Delta_h$. To close the bulk gap of SbAsX$_2$, we chose elastic strain, an effective strategy[41-43] for engineering the electronic properties of 2D materials, to tune the strength of SOC and then change the gap. Fig. 2(f) plots the variation of band gap (with SOC) as a function of in-plane biaxial tensile strain for SbAsX$_2$ MLs. Under the increasing strain, a consistent trend is observed for different X — the band gap is gradually decreased to zero at a critical strain and then increases. The critical strain for gap closure is nearly proportional to the electronegativity of functional group X [Fig. 2(g)]. Since H atom has the lowest electronegativity among the considered functional groups, SbAsH$_2$ holds the smallest critical strain of 2.3%, which is rather accessible in experiment. Hereafter, our further discussion focuses on SbAsH$_2$ only, whereas the other SbAsX$_2$ MLs share the similar behaviors but require much larger critical strain.

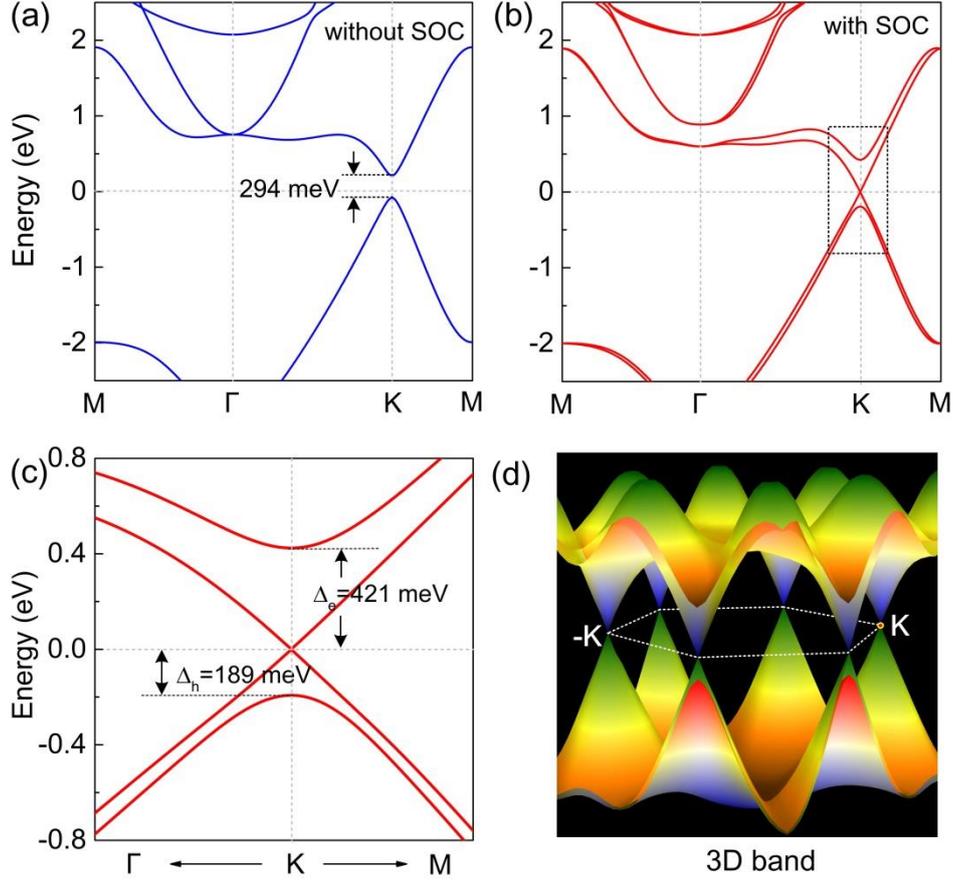

FIG. 3. Electronic band structures of 2.3-SbAsH$_2$ ML (a) without SOC and (b) with SOC. (c) Enlarged plot for the band structure with SOC near the Fermi level around the K point. (d) 3D Dirac cones and their nearest neighbor valence and conduction bands. The Dirac state has been confirmed by HSE06 calculations (see Fig. S6 in the Supporting Information). The band structures of other SbAsX$_2$ (X=F, Cl, Br, I) under the critical strain are presented in Fig. S7. The more accurate values of spin-splitting of $\Delta_e$ and $\Delta_h$ from HSE06 calculations are listed in Table I.

Fig. 3 depicts the band structure of SbAsH$_2$ at the critical tensile strain of 2.3% (termed as 2.3-SbAsH$_2$ thereafter). After considering SOC, the original direct semiconductor [Fig. 3(a)] closes its band gap and forms six spin-splitting Dirac cones with linear dispersion at the corners of BZ on the Fermi level [Figs. 3(b, c, d)]. Since the Dirac state is spin-non-degenerate, the Dirac points are actually Weyl points. The maximal spin-splitting at the K point is as large as 421 meV ($\Delta_e$) and 189 meV ($\Delta_h$) for the lowermost conduction band (LCB) and uppermost valence band (UVB), respectively, which is one order of magnitude greater than the thermal energy (25 meV) at room temperature. This is highly desirable for avoiding spin-flip scattering in spintronics applications.[44,45] More importantly, only linear Dirac bands emerge in such giant spin-splitting gap ($\Delta_e+\Delta_h$), which means that all the charge carriers would be massless if the Fermi level locates in this gap window. The Fermi velocity $v_F$ (see Table I) of these massless carriers can be evaluated using a linear fitting: $\hbar v_F \approx dE(\bm{k})/d\bm{k}$. Remarkably, the obtained $v_F = 8.79\times10^5$ m/s for SbAsH$_2$ is even comparable with that of graphene ($9.42\times10^5$ m/s[46]).

Having confirmed that the Dirac state of 2.3-SbAsH$_2$ is spin-splitting, it is necessary to further examine the orientation of spin moments. Fig. 4(a) presents the LCB and UVB of 2.3-SbAsH$_2$ along $\bm{k}$-path of Γ-K-M-(−K)-Γ with the projection of spin operator $\hat{s}_z$, i.e., $\langle\psi_{n'\bm{k}}|\hat{s}_z|\psi_{n\bm{k}}\rangle$. It can be seen that the spin-splitting Dirac states at the K and −K valleys have opposite spin moments. This means that the low-energy Dirac fermions in the K and −K valleys can be distinguished by their spin index. Once the K and −K valleys are separated in transport, 100% out-of-plane spin polarization would be realized in such system. Additionally, breaking of the inversion symmetry will make the Dirac fermions acquiring a valley-contrasting Berry curvature $\Omega^z(\bm{k})$, which can be calculated from the Kubo formula:[47-49]

$$\Omega^z(\bm{k})=\sum_n f_n \Omega_n^z(\bm{k}), \quad (1)$$

where $f_n$ is the Fermi-Dirac distribution, and $\Omega_n^z(\bm{k})$ can be obtained from

$$\Omega_n^z(\bm{k}) = -\sum_{n' \neq n} \frac{2\,\mathrm{Im}\,\langle \psi_{n\bm{k}} | \bm{v}_x | \psi_{n'\bm{k}} \rangle \langle \psi_{n'\bm{k}} | \bm{v}_y | \psi_{n\bm{k}} \rangle}{(E_{n'} - E_n)^2}. \tag{2}$$

Here $|\psi_{n\bm{k}}\rangle$ is the Bloch state with the eigenvalue $E_n$, and $\bm{v}_{x(y)}$ are the velocity operators. From the construction of maximally localized Wannier functions, the Berry curvature of the occupied states for 2.3-SbAsH$_2$ can be obtained, as shown in Fig. 4(b). $\Omega^z(\bm{k})$ exhibits obvious peaks at both K and −K valleys but with opposite sign;

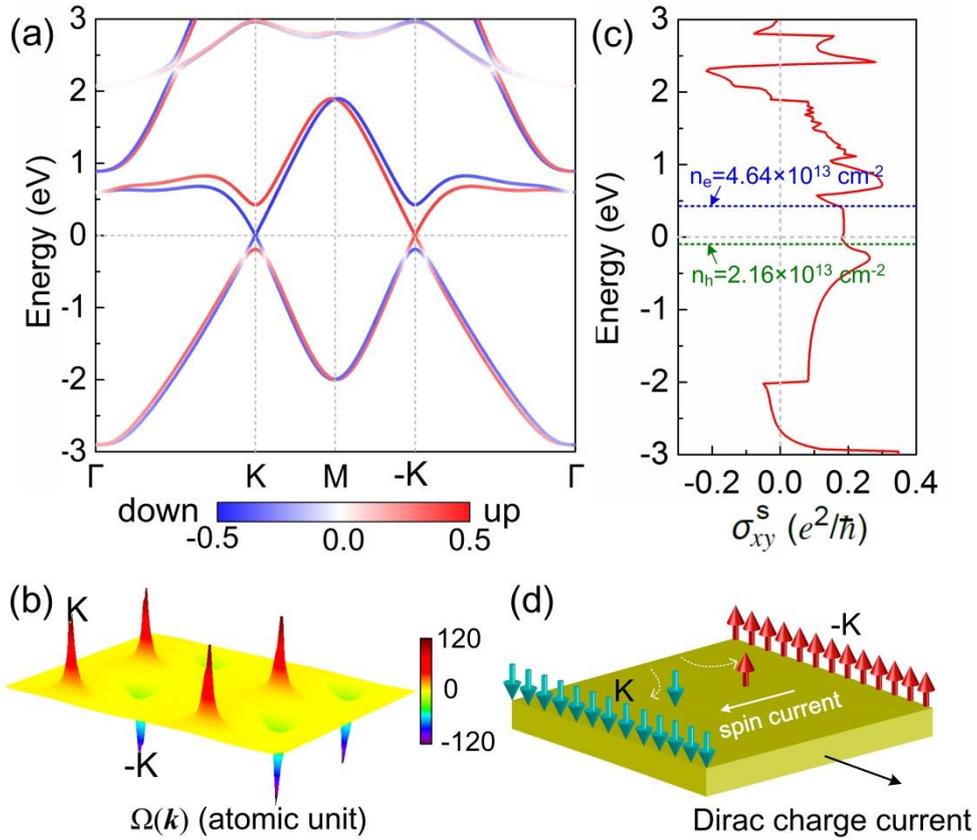

FIG. 4. (a) Color map of the band structure of 2.3-SbAsH$_2$ with the projection of spin operator $\hat{s}_z$. (b) Berry curvatures of 2.3-SbAsH$_2$ in the $k_z$=0 plane. (c) Intrinsic spin Hall conductivity $\sigma^s_{xy}$ (e$^2$/ℏ) as a function of Fermi energy for 2.3-SbAsH$_2$. The dashed lines indicate the positions of the Fermi level with carrier concentration of $4.64 \times 10^{13}$ cm$^{-2}$ for p-doped sample and $2.16 \times 10^{13}$ cm$^{-2}$ for n-doped sample, respectively. (d) Schematic Dirac spin-valley Hall effect in svc-DSMs for generating spin current and spin-valley polarized Dirac fermions.

consequently the distribution of $\Omega^z(\boldsymbol{k})$ exhibits $C_{3v}$ symmetry. In other words, one cannot distinguish Dirac fermions in these two kinds of valleys from energy; but it is feasible to differentiate them by their opposite Berry curvatures as well as their opposite out-of-plane spin moments.

When an external in-plane electric field is applied, the valley-contrasting Berry curvatures would induce two opposite anomalous velocities[50]

$$\dot{\boldsymbol{x}} = \frac{1}{\hbar}\frac{\partial E_n(\boldsymbol{k})}{\partial \boldsymbol{k}} + \dot{\boldsymbol{k}} \times \Omega^z(\boldsymbol{k}) \qquad (3)$$

for the motion of charge carriers, which is perpendicular to the field. Since the valley index is coupled with spin in 2.3-SbAsH$_2$, the valley Hall and spin Hall effects would occur simultaneously. Namely, the applied in-plane electric field would lead the charge carriers of different valleys to flow to the opposite transverse edges, resulting in both valley and spin polarization along the edges [see Fig. 4(d)]. Because the charge carries of 2.3-SbAsH$_2$ are massless Dirac fermions, here we term such Hall effect as *Dirac spin-valley Hall effect*. Note that the Dirac valley index is defined only within the energy windows of spin-splitting $\Delta_e$ for LCB and $\Delta_h$ for UVB [see Fig. 3(c)]. Thus, the valley Hall conductivity should coincide with the spin Hall conductivity if the Fermi level lies in the $\Delta_e$ or $\Delta_h$ window.

To quantitatively examine the transverse anomalous motion in the proposed Dirac spin-valley Hall effect, we calculate the intrinsic spin Hall conductivity (ISHC) under zero temperature and clean limit, which is given by

$$\sigma_{xy}^s = \frac{e}{\hbar} \int_{V_G} \frac{\mathrm{d}^2\boldsymbol{k}}{(2\pi)^2} \Omega^{zs}(\boldsymbol{k}). \qquad (4)$$

Here, the spin Berry curvature $\Omega^{zs}(\boldsymbol{k})$ can be calculated again using the Kubo formula[47-49] like the ordinary $\Omega^{z}(\boldsymbol{k})$. Fig. 4(c) illustrates $\sigma_{xy}^{s}$ as a function of Fermi level for 2.3-SbAsH$_2$ ML. Overall speaking, $\sigma_{xy}^{s}$ displays a relatively complicated behavior in the considered energy window (−3 ~ 3 eV), undergoing both numerical oscillations and sign changes that can be used to distinguish the intrinsic spin Hall effect from the extrinsic one.[51] Since the highest carrier concentration attained for graphene is about $10^{13}$ cm$^{-2}$ in experiment,[1] it is essential to focus on the situation of $\sigma_{xy}^{s}$ in a realizable doping range from −0.16 eV ($n_h$=2.16×10$^{13}$ cm$^{-2}$) to 0.39 eV ($n_e$=4.64×10$^{13}$ cm$^{-2}$) [see the dashed lines in Fig. 4(c)]. Remarkably, $\sigma_{xy}^{s}$ nearly keeps a constant value (~0.19 e$^2$/ℏ) in this range. As a matter of fact, such energy range just locates in the large spin-splitting gap $\Delta_h+\Delta_e$, in which the bands exhibit linear energy dispersion without extraneous non-Dirac bands. Therefore, the clean Dirac state in the spin-splitting gap should mainly account for the constant $\sigma_{xy}^{s}$.

According to the classification of possible gap closure in 2D systems,[52,53] one may conjecture that a topological phase transition would occur in SbAsH$_2$ during the variation of band gap [see Fig. 2(f)]. To explore this, we first calculate the orbital-resolved band structures in the vicinity of the Fermi level (see Fig. S8 in the Supporting Information). The results reveal a band inversion between $|Sb, p_{xy}\rangle + |As, s\rangle$ and $|As, p_{xy}\rangle + |Sb, s\rangle$ induced by tensile strain. Then, we calculate the Wannier charge center (WCC) and the edge states for SbAsH$_2$ with 0% and 5% tensile strains. Taking an arbitrary horizontal line [e.g., WCC=0.4 in Figs. 5(a) and 5(b)] as reference, one can see that the number of crossings between the reference line and the evolution of WCC is even for the strain-free $SbAsH_2$, and it is odd for $SbAsH_2$ under 5% strain. This means that the pristine $SbAsH_2$ is a trivial insulator with $\mathbb{Z}_2=0$, while 5% strained $SbAsH_2$ is a topological insulator with $\mathbb{Z}_2=1$. As shown in Fig. 5(c), the edge states start from the conduction band and then re-enter the conduction band, further evidencing that the strain-free $SbAsH_2$ is a trivial insulator. In contrary, for the $SbAsH_2$ under 5% strain, there is a pair of gapless non-trivial edge states (*i.e.*, spin transport channels) traversing across the bulk gap and connecting the conduction and valence bands.

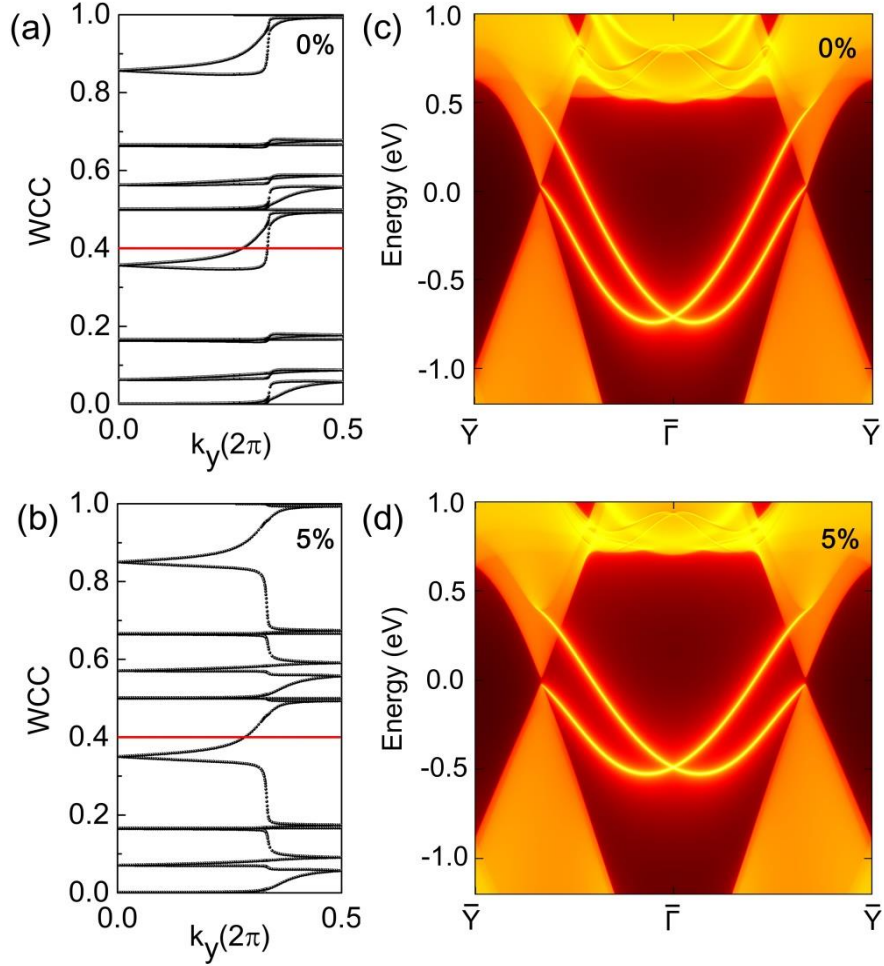

FIG. 5. Evolution of the Wannier charge centers (WCCs) along $k_y$ for (a) strain-free and (b) 5% strained $SbAsH_2$; the edge states of semi-infinite (c) $SbAsH_2$ and (d) that under 5% tensile strain.

Clearly, the confirmed svc-DSM state of 2.3-SbAsH$_2$ occurs at the boundary between trivial and topological insulators, similar to the 3D Dirac semimetallic state that is located at the topological phase boundary and can be driven into various topologically distinct phases.[54-56] Unlike the 3D case, however, the 2D svc-DSM state of 2.3-SbAsH$_2$ is not protected by nonsymmorphic space group symmetry. Notwithstanding this, applying strain to a 2D material is rather controllable in experiment (e.g., by stretching or bending of a flexible substrate on which the 2D material is attached).[57-59] In this sense, one can not only tune the SbAsH$_2$ ML into the desired svc-DSM state by modest strain, but also readily realize the reversible transition between the trivial and topological states.

In summary, our *ab initio* calculations demonstrate that a class of svc-DSMs with spin and valley polarized Dirac state can emerge in the functionalized SbAs MLs under accessible strain. With inversion symmetry breaking, giant spin splitting is induced by significant SOC effect at the corners of BZ, giving rise to a clean Dirac state in a wide energy window that is far larger than the thermal energy at room temperature. Excitingly, the craved massless and dissipationless quantum transport can be realized within such energy window based on the Dirac spin-valley Hall effect, which is manifested by the constant spin Hall conductivity. Our topological analysis shows that such svc-DSM occurs at the boundary between trivial and topological insulators. These results provide a promising platform for the fantastic integration of Dirac physics, spintronics and valleytronics, offering new opportunities for the realization of ultra-fast and ultra-low-power nanodevices. In view of the recent experimental progress in 2D group VA materials,[60-62] Janus monolayers[63,64] and functionalized 2D materials,[38,39,65] our findings should facilitate further exploration for the intriguing svc-DSMs. A major challenge for the subsequent studies is to find the intrinsic svc-DSMs that are topologically protected by symmetry. The potential candidates might be the 2D nonsymmorphic crystal with inversion asymmetry and providential SOC strength.

## ASSOCIATED CONTENT

**Supporting Information**

The Supporting Information is available free of charge on the *** Publications website at DOI: ****.

The mechanical exfoliation, structure and stability of SbAs ML; the band structures of SbAs and $SbAsX_2$ (X= H, F, Cl, Br and I) MLs; band structure of 2.3-$SbAsH_2$ ML from HSE06 calculations; svc-DSM states of $SbAsX_2$ (X=F, Cl, Br and I) MLs; strain induced band inversion in 2.3-$SbAsH_2$ ML.

## AUTHOR INFORMATION

**Corresponding Author**

*E-mail: zhaojj@dlut.edu.cn

**Notes**

The authors declare no competing financial interest.

**ACKNOWLEDGMENTS**

We acknowledge useful discussions with Prof. S.-W. Gao. This work is currently supported by the National Natural Science Foundation of China (11604165, 11574040, 11874085, 11374033) and Natural Science Foundation of Inner Mongolia (2016BS0104). W.F. also acknowledges the funding through an Alexander von Humboldt Fellowship. The Supercomputing Center of Dalian University of Technology is acknowledged for providing computing resources.

(56) Wang, Z.; Sun, Y.; Chen, X.-Q.; Franchini, C.; Xu, G.; Weng, H.; Dai, X.; Fang, Z. Dirac semimetal and topological phase transitions in $A_3Bi$ (A=Na, K, Rb). *Phys. Rev. B* **2012**, *85*, 195320.

(57) Guinea, F.; Katsnelson, M. I.; Geim, A. K. Energy gaps and a zero-field quantum Hall effect in graphene by strain engineering. *Nat. Phys.* **2009**, *6*, 30.

(58) Akinwande, D.; Petrone, N.; Hone, J. Two-dimensional flexible nanoelectronics. *Nat. Commun.* **2014**, *5*, 5678.

(59) Fei, R.; Yang, L. Strain-Engineering the Anisotropic Electrical Conductance of Few-Layer Black Phosphorus. *Nano Lett.* **2014**, *14*, 2884-2889.

(60) Zhang, S.; Guo, S.; Chen, Z.; Wang, Y.; Gao, H.; Gomez-Herrero, J.; Ares, P.; Zamora, F.; Zhu, Z.; Zeng, H. Recent progress in 2D group-VA semiconductors: from theory to experiment. *Chem. Soc. Rev.* **2018**, *47*, 982-1021

(61) Reis, F.; Li, G.; Dudy, L.; Bauernfeind, M.; Glass, S.; Hanke, W.; Thomale, R.; Schäfer, J.; Claessen, R. Bismuthene on a SiC substrate: A candidate for a high-temperature quantum spin Hall material. *Science* **2017**, *357*, 287-290.

(62) Gusmão, R.; Sofer, Z.; Bouša, D.; Pumera, M. Pnictogen (As, Sb, Bi) Nanosheets for Electrochemical Applications Are Produced by Shear Exfoliation Using Kitchen Blenders. *Angew. Chem. Int. Edit.* **2017**, *129*, 14609-14614.

(63) Lu, A.-Y.; Zhu, H.; Xiao, J.; Chuu, C.-P.; Han, Y.; Chiu, M.-H.; Cheng, C.-C.; Yang, C.-W.; Wei, K.-H.; Yang, Y.; Wang, Y.; Sokaras, D.; Nordlund, D.; Yang, P.; Muller, D. A.; Chou, M.-Y.; Zhang, X.; Li, L.-J. Janus monolayers of transition metal dichalcogenides. *Nat. Nanotech.* **2017**, *12*, 744.

(64) Zhang, J.; Jia, S.; Kholmanov, I.; Dong, L.; Er, D.; Chen, W.; Guo, H.; Jin, Z.; Shenoy, V. B.; Shi, L.; Lou, J. Janus Monolayer Transition-Metal Dichalcogenides. *ACS Nano* **2017**, *11*, 8192-8198.

(65) Tucek, J.; Hola, K.; Bourlinos, A. B.; Blonski, P.; Bakandritsos, A.; Ugolotti, J.; Dubecky, M.; Karlicky, F.; Ranc, V.; Cepe, K.; Otyepka, M.; Zboril, R. Room temperature organic magnets derived from sp3 functionalized graphene. *Nat. Commun.* **2017**, *8*, 14525.
24

# Supporting Information for
# Two-dimensional spin-valley-coupled Dirac semimetals in functionalized SbAs monolayers

Zhifeng Liu,[†‡] Wangxiang Feng,[§∥] Hongli Xin,[†] Yinlu Gao,[⊥] Pengfei Liu,[‡#] Yugui Yao,[§] Hongming Weng,[£¶] Jijun Zhao[‡⊥*]

[†]*School of Physical Science and Technology, Inner Mongolia University, Hohhot 010021, China*

[‡]*Beijing Computational Science Research Center, Beijing 100094, China*

[§]*School of Physics, Beijing Institute of Technology, Beijing 100081, China*

[∥]*Peter Grünberg Institut and Institute for Advanced Simulation, Forschungszentrum Jülich and JARA, 52425 Jülich, Germany*

[⊥]*Key Laboratory of Materials Modification by Laser, Ion and Electron Beams (Dalian University of Technology), Ministry of Education, Dalian 116024, China*

[#]*Institute of High Energy Physics, Chinese Academy of Sciences (CAS), Beijing 100049, China*

[£]*Beijing National Laboratory for Condensed Matter Physics and Institute of Physics, Chinese Academy of Sciences, Beijing 100190, China*

[¶]*Songshan Lake Materials Laboratory, Guangdong 523808, China*

I.   Mechanical exfoliation for SbAs ML

II.  Structure and stability of freestanding SbAs ML

III. Band structures of SbAs ML

IV.  Band structures of SbAsH$_2$ ML

V.   Band structures of SbAsX$_2$ (X=F, Cl, Br and I) MLs

VI.  Band structure of 2.3-SbAsH$_2$ ML from HSE06

VII. svc-DSM states of SbAsX$_2$ (X=F, Cl, Br and I) MLs

VIII. Strain induced band inversion

---

[*] Corresponding author. Email: zhaojj@dlut.edu.cn

## I. Mechanical exfoliation for SbAs ML

Since the layered structure of bulk SbAs [see Figs. S1(a) and S1(b)] has been synthesized in experiment, we calculate the cleavage energy by introducing a fracture in it so as to examine whether the SbAs ML can be obtained from the bulk sample via mechanical exfoliation in experiment. As shown in Fig. S1(c), the calculated cleavage energy quickly increases with the increasing separation distance ($d-d_0$) and then saturates to a value corresponding to the exfoliation energy about 0.34 J/m$^2$. For comparison, the cleavage energies of some representative layered materials (e.g., graphite, h-BN and MoS$_2$) have also been calculated [see Fig. S1(d)]. Remarkably, the 0.34 J/m$^2$ exfoliation energy of SbAs is smaller than that of the famous graphite (0.38 J/m$^2$) and h-BN (~0.51 J/m$^2$), suggesting that it is feasible to obtain the SbAs ML by mechanical exfoliation from the existing bulk like the graphene. It should be noted that the reliability of our calculated results can be confirmed by the agreement between our results and the previous data, such as 0.38 J/m$^2$ versus 0.37 J/m$^2$ [1] for graphite, and 0.29 J/m$^2$ versus 0.27 J/m$^2$ [2] for MoS$_2$.

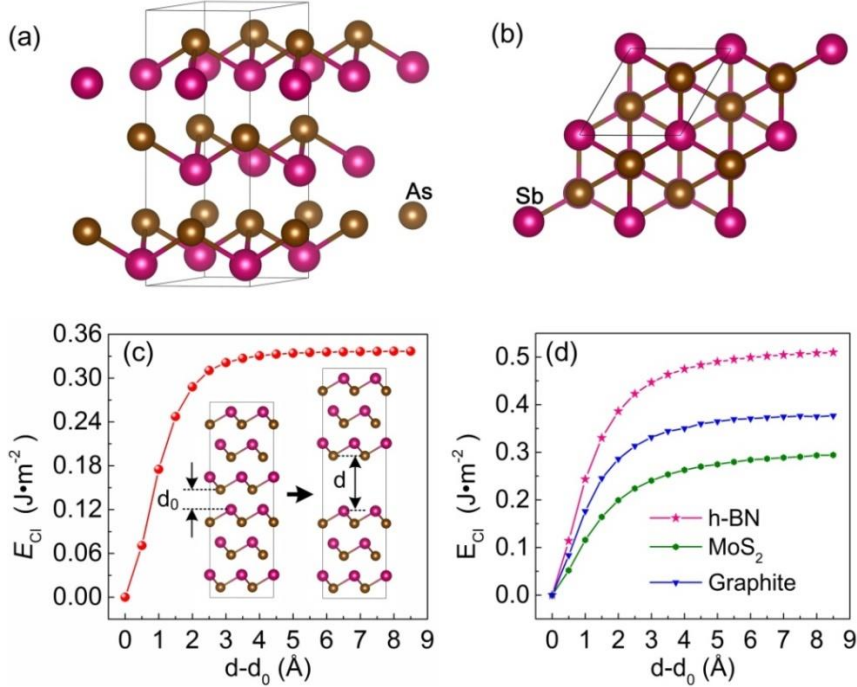

FIG. S1. (a) Side and (b) top view of the optimized atomic structure for the chemical ordered SbAs solid. (c) Cleavage energy $E_{cl}$ as a function of the separation distance $d-d_0$ between two fractured parts of layered SbAs. For comparison, the corresponding (d) cleavage energy of layered $h$-BN, MoS$_2$ and graphite are also presented.

## II. Structure and stability of freestanding SbAs ML

As shown in Figs. S2(a) and S2(b), the optimized atomic structure of pristine SbAs ML is buckled as the case of silicene [3], which derives from the relatively weak bonding between Sb and As atoms. Such buckling is benefit to enhance the overlap between $\pi$ and $\sigma$ orbitals and stabilizes the system. Its dynamical and thermal stability has been respectively confirmed by our phonon calculation that shows non imaginary frequencies [Fig. S2(c)], and *ab initio* molecular dynamic (AIMD) simulation in which there are no structural destruction [Fig. S2(d)].

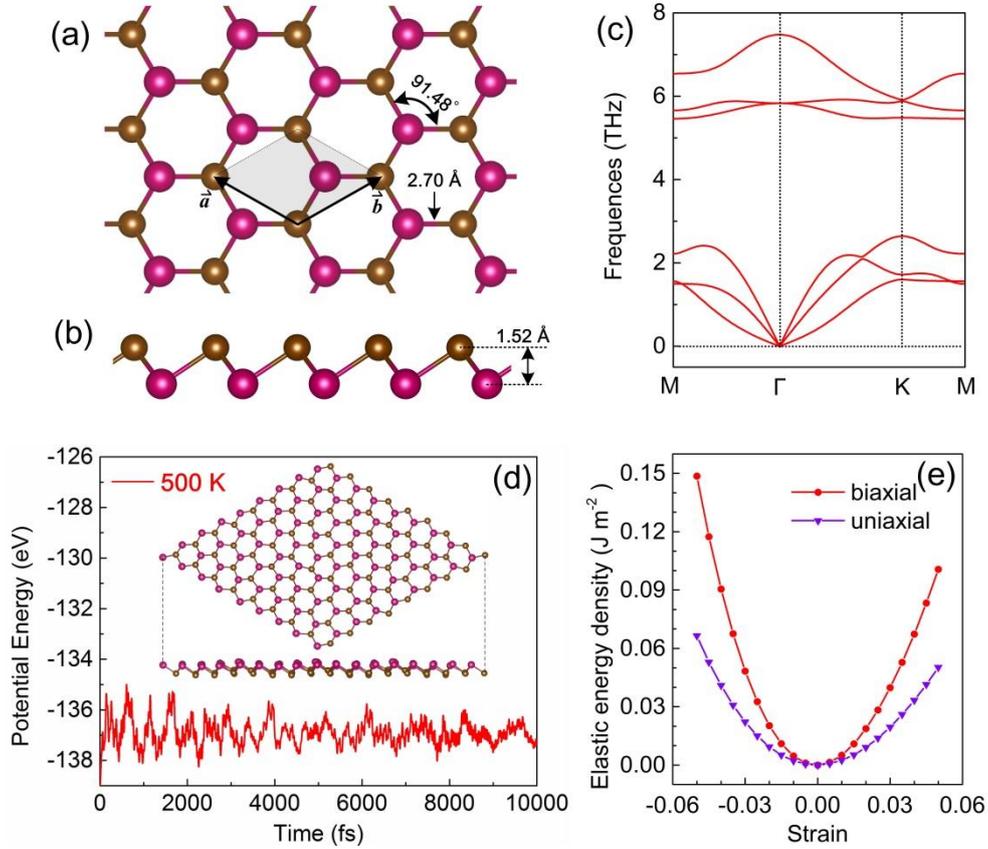

FIG. S2. Top (a) and side (b) views of optimized structure, phonon dispersion (c), total potential energy fluctuation (d) during the AIMD simulations at 500K, and strain energies under different biaxial and uniaxial strains for SbAs ML. The inset in (d) shows the snapshot at the end of simulation of 10 ps.

In experiment, the chemical functionalization of SbAs ML requires that the exfoliated SbAs ML should be able to withstand its own weight and form a freestanding membrane. To check this aspect, we calculate the in-plane stiffness of the

SbAs ML characterized by in-plane Young's modulus $Y_{2D}$, which is defined as

$$Y_{2D} = \frac{C_{11}^2 - C_{12}^2}{C_{11}}, \tag{1}$$

where $C_{11}$ and $C_{12}$ is the corresponding linear elastic constants. For 2D materials, the elastic energy density $U(\varepsilon)$ can be expressed as [4]:

$$U(\varepsilon) = \frac{1}{2}C_{11}\varepsilon_{xx}^2 + \frac{1}{2}C_{22}\varepsilon_{yy}^2 + C_{12}\varepsilon_{xx}\varepsilon_{yy} + 2C_{44}\varepsilon_{xy}^2. \tag{2}$$

Thus, the elastic constants of $C_{11}$, $C_{12}$ can be derived by fitting the curves of $U(\varepsilon)$ with respect to uniaxial and equi-biaxial strain [Fig. S2 (e)]. The $Y_{2D}$ of SbAs ML is calculated to be 46.4 GPa·nm. On the basis of the calculated $Y_{2D}$, the gravity induced typical out-of-plane deformation $h$ can be estimated by [5]

$$\frac{h}{l} = \left(\frac{\sigma g l}{Y_{2D}}\right)^{1/3}, \tag{3}$$

where $\sigma$ (=2.53×10$^{-6}$ Kg/m$^2$) is the surface density of SbAs ML, $l$ is the size of possible free-standing SbAs flakes. If we assume a macroscopic size $l \approx 100$ μm, the $h/l$ is calculated to be 3.76×10$^{-4}$. This value is of the same order of magnitude as that of graphene [5]. This indicates that without the support of a substrate, the in-plane stiffness of SbAs ML is strong enough to withstand its own weight and keep its free-standing membrane.

## III. Band structure of SbAs ML

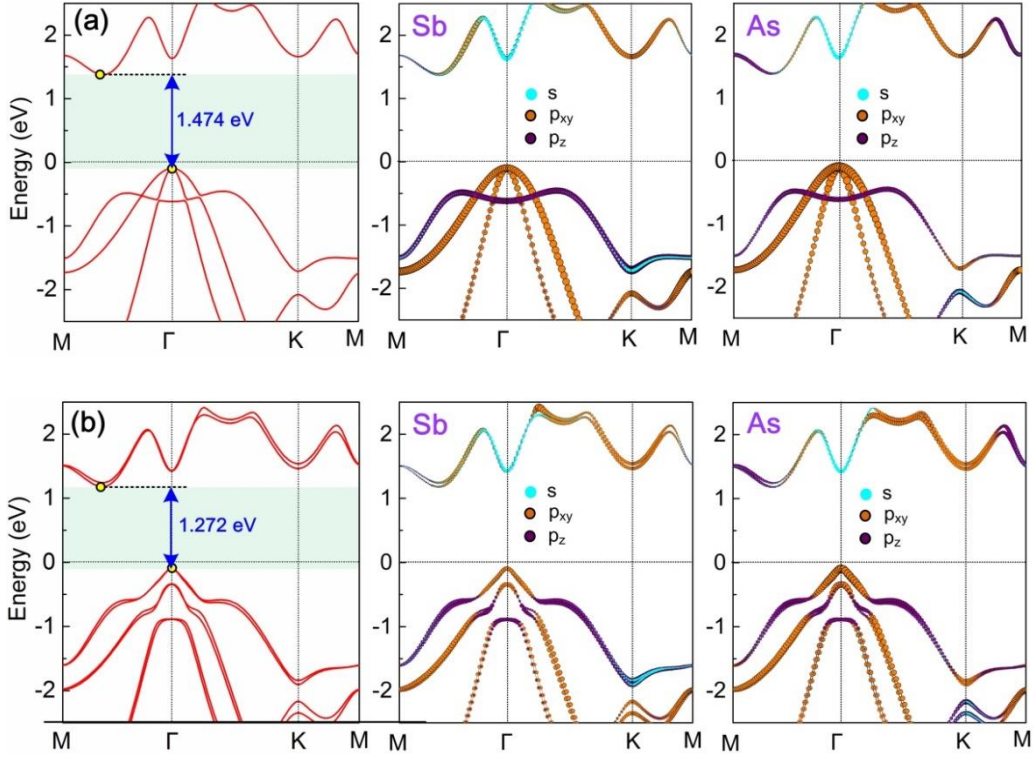

FIG. S3. Electronic band structures of SbAs ML (a) without SOC and (b) with SOC. The corresponding orbital-resolved electronic band structures for Sb and As atoms are displayed at the same row. Clearly, SbAs ML is an indirect semiconductor. The valence band maximum (VBM) at Γ point is mainly contributed by hybridization of Sb-$p_{xy}$ and As-$p_{xy}$ states, while the conduction band minimum (CBM) at the originates from Sb-$p_z$ and As-$p_z$ states, as well as few contribution of $s$ state. Note that at the higher-lying valence band, the contribution of $p_z$ for both Sb and As is obvious. When the SOC effect is considered, the spin-splitting at CBM leads to a distinct decrease in band bap (from 1.474 eV to 1.272 eV).

## IV. Band structure of SbAsH$_2$ ML

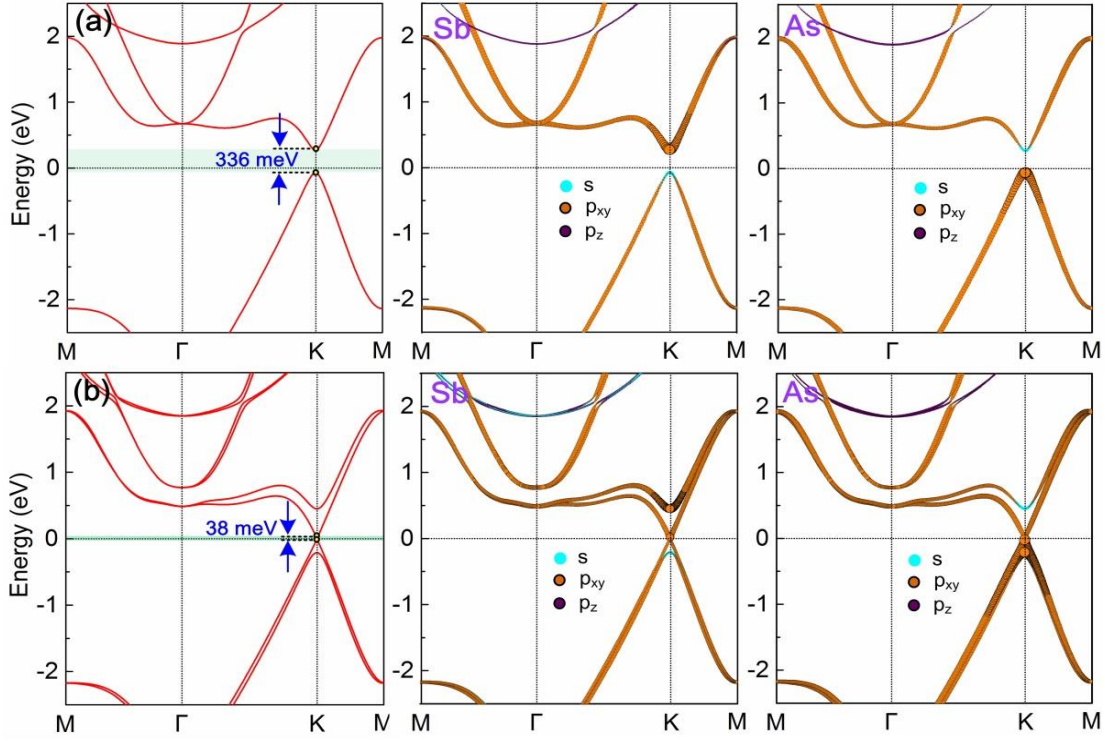

FIG. S4. Electronic band structures of SbAsH$_2$ ML (a) without SOC and (b) with SOC. The corresponding orbital-resolved electronic band structures for Sb and As atoms are displayed at the same row. Unlike the case of SbAs ML, SbAsH$_2$ is a direct semiconductor with greatly reduced band gap at K point. Since the electronegativity of H atom is larger than that of both Sb and As atoms, the $p_z$ electrons of the SbAs ML transfer to the functional group H atoms. Thus the Fermi level moves down, and the $p_z$ states of SbAs ML strongly couple with the H-$s$ state at the deep energy level. As shown in the figure, the $p_z$ state is almost no contribution to both valance and conduction bands in the vicinity of new Fermi level for SbAsH$_2$. The VBM is mainly contributed by Sb-$p_{xy}$ and As-$s$ orbitals, while the main contribution of CBM is Sb-s and As-$p_{xy}$ orbitals. Interestingly, in the present of SOC, giant spin-splitting can be found at the inversion asymmetric K point, which leads to a further reduced band gap (38 meV).

## V. Band structures of SbAsX$_2$ (X=F, Cl, Br and I) ML

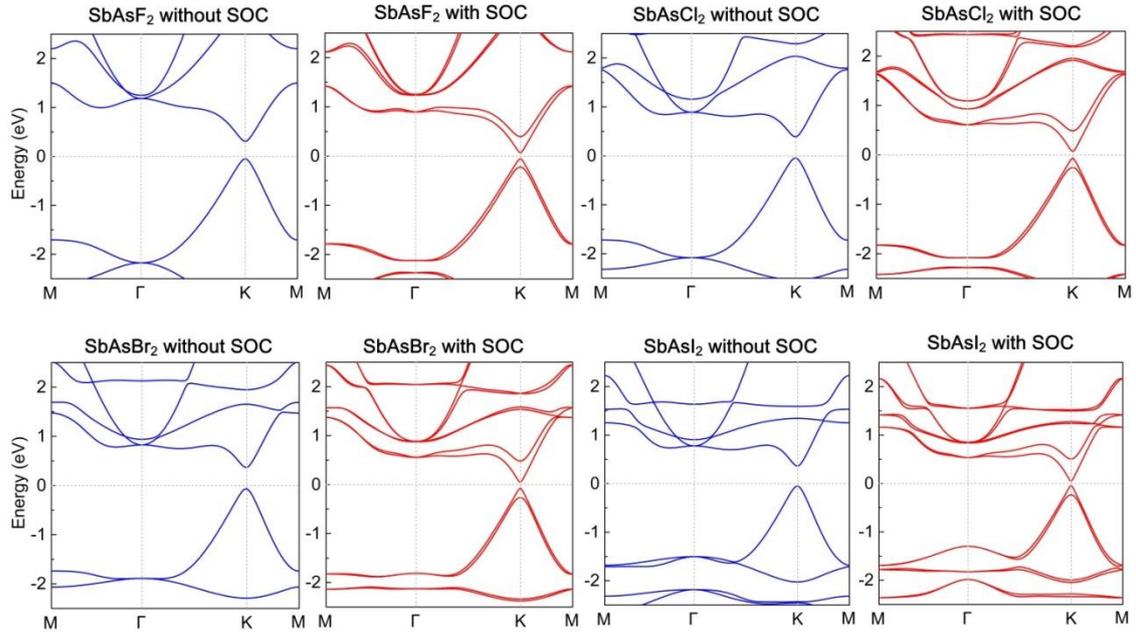

FIG. S5. Electronic band structures of SbAsX$_2$ (X=F, Cl, Br and I) without SOC and with SOC. The distribution of orbitals is similar with the case of SbAsH$_2$ (see Fig. S4)

## VI. Band structure of 2.3-SbAsH$_2$ ML from HSE06

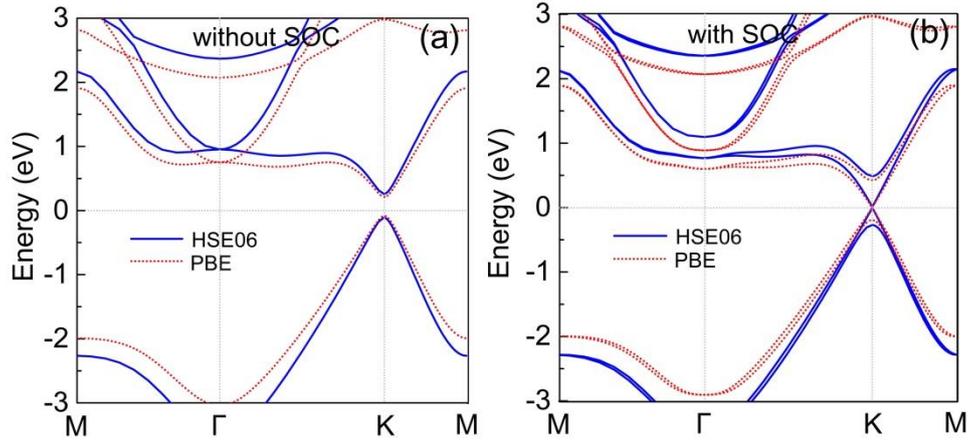

FIG. S6. The band structure (blue solid line) calculated from HSE06 method (a) without SOC and (b) with SOC for the 2.3-SbAsH$_2$ ML. For comparison, the bands at PBE theory level are also displayed (red dashed line). Although the positions of most bands show relative shifts in these two methods, those of VBM and CBM at K point stay still. Thus, the spin-splitting Dirac state is well preserved, confirming that the 2.3-SbAsH$_2$ is indeed a Dirac semimetal.

## VII. svc-DSM state of SbAsX$_2$ (X=F, Cl, Br and I) MLs

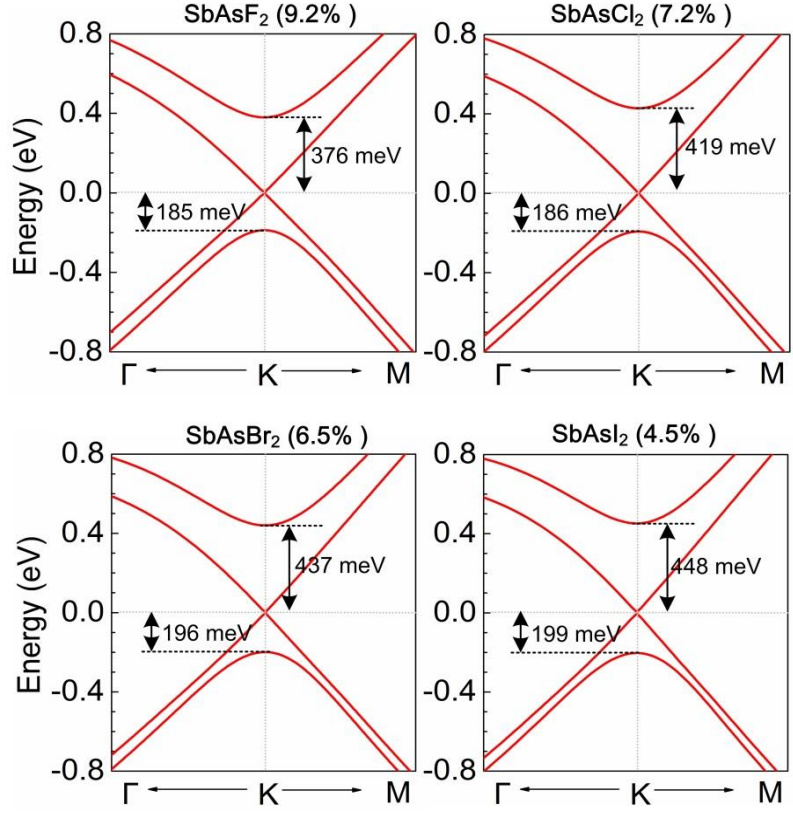

FIG. S7. The enlarged band structure with SOC near the Fermi level around the K point for SbAsX$_2$ (X=F, Cl, Br and I) MLs under critical strain. Clearly, similar to SbAsH$_2$, the svc-DSM state is formed in the critical strain for all the other SbAsX$_2$ MLs.

## VIII. Strain induced band inversion

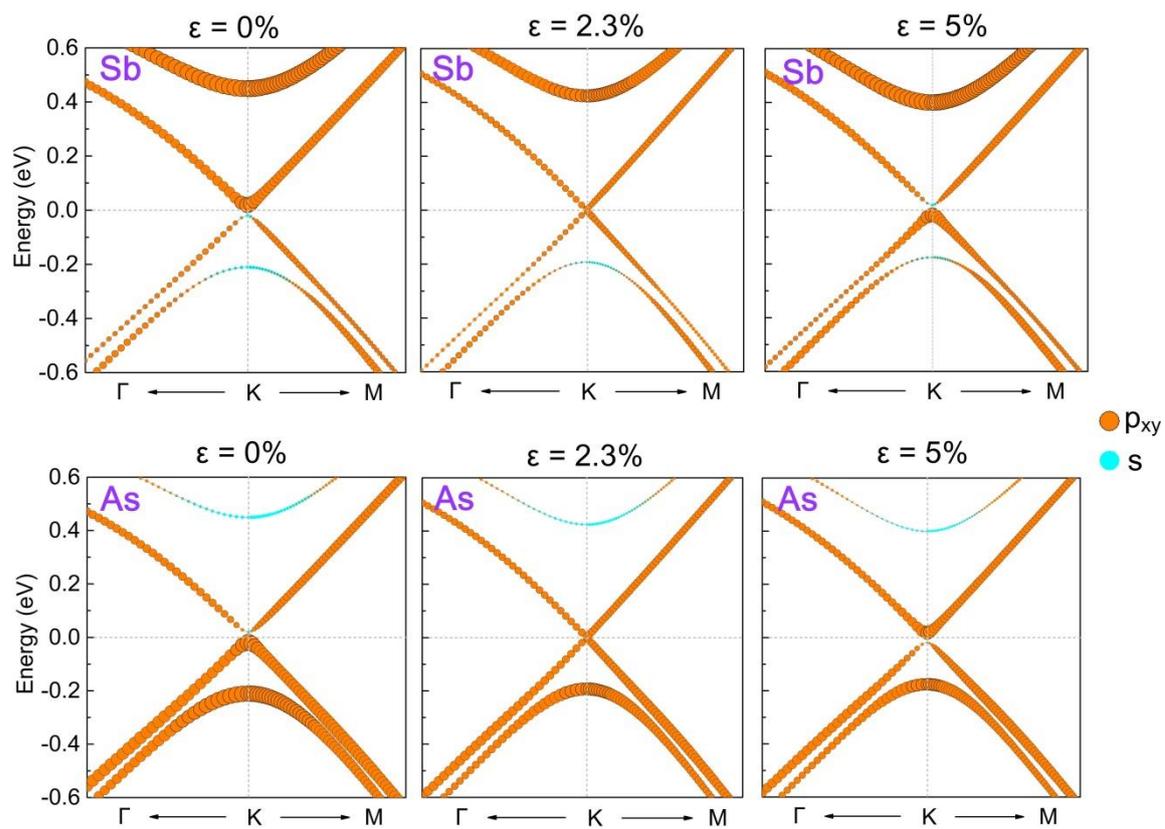

FIG. S8. The orbital-resolved band structures in the vicinity of Fermi level for Sb and As atoms of SbAsH$_2$ ML with zero, 2.3% and 5% tensile strain.